\documentclass[twocolumn,showpacs]{revtex4}
\usepackage{dcolumn,amssymb}
\usepackage{amsmath,epsfig}
\usepackage{rotating}

\newcommand{\C}{\mathbb{C}}

\newcommand{\R}{\mathbb{R}}

\newcommand{\cP}{{\cal P}}

\newcommand{\bea}{\begin{eqnarray}}
\newcommand{\eea}{\end{eqnarray}}

\newcommand{\tr}{{\rm tr}}
\newcommand{\one}{\mbox{1\hspace{-.8ex}I}}
\def\be#1\ee{\begin{equation}#1\end{equation}}

\newcommand{\bra}{\langle}
\newcommand{\ket}{\rangle}

\def\be#1\ee{\begin{equation}#1\end{equation}}

\begin{document}

\title
{Gauge cooling in complex Langevin for QCD with heavy quarks}

\author{ Erhard Seiler$^1$}\email{ehs@mppmu.mpg.de}
\author {D{\'e}nes Sexty$^2$}\email{d.sexty@thphys.uni-heidelberg.de}
\author{Ion-Olimpiu Stamatescu$^2$} 
\email{I.O.Stamatescu@thphys.uni-heidelberg.de}
\affiliation{$^1$Max-Planck-Institut f\"ur Physik
(Werner-Heisenberg-Institut), M\"unchen, Germany}
\affiliation{$^2$Institut f\"ur Theoretische Physik, Universit\"at Heidelberg, 
Germany}
\date{\today}
\begin{abstract}
\noindent
We employ a new method, ``gauge cooling'', to stabilize complex Langevin 
simulations of QCD with heavy quarks. The results are checked against 
results obtained with reweigthing; we find agreement within the estimated 
errors.  The method allows us to go to previously unaccessible high 
densities. 
\end{abstract}
\maketitle

{\it Introduction.} -- An important problem of particle physics is to 
derive the phase diagram of hot and dense QCD from first principles. The 
difficulty lies in the fact that the action becomes complex at finite 
density. Various methods have been employed to get around this problem: 
Taylor expansions at $\mu=0$ and analytic continuation from imaginary 
$\mu$, reweighting etc., but these have a limited range of applicability 
\cite{pdf}. A general method that recently has received a lot of attention 
is the Complex Langevin Equation (CLE)  \cite{bergesstam, bergessexty, 
guralnik, aarts, aj, aartsstam, ajss} . It has had some successes as well 
as some failures; the failures seem to be related to the problem that the 
system wants to spend ``too much time too far out'' in the complexified 
configuration space; this problem and possible ways to deal with it were 
discussed in \cite{ajss,opt}.

In QCD at finite density the enlarged gauge freedom in the complexified 
field space can actually open a new way to limit the dangerous large 
excursions by employing non-unitary gauge transformations. This freedom 
was used in a somehwat different way for a $SU(2)$ toy model \cite{opt}; 
the procedure used there, however, does not generalize to a lattice model 
in an obvious way.

Here we introduce the method of gauge cooling (g.c.), designed to keep the 
link variables as close as possible to the unitaries, thereby preventing 
the dangerous large excursions.  We test the idea first for a simple 
Polyakov loop model, where we have exact results to compare our data with; 
then we apply it to a heavy quark approximation of QCD with chemical 
potential (HQCD) (cf. \cite{aartsstam}).

{\it Complex Langevin for gauge models} -- The complex Langevin method 
\cite{parisi,klauder} for a complex action $S$ is based on setting 
up a stochastic process on the {\it complexification} of the configuration 
space. The longtime average of holomorphic observables is then supposed to 
give the correct average corresponding to the complex weight $\exp(-S)$. 
In lattice models of QCD the configuration space is $SU(3)^{\# {\rm 
links}}$ (see \cite{batrouni} for the Real Langevin approach);
 after complexification this becomes $SL(3,\C)^{\# {\rm links}}$ 
\cite{gausterer}. The complex Langevin equation (CLE) is
\be 
(dU_{x,\mu}) U_{x,\mu}^{-1} = -\sum_a\lambda_a (D_{a,x,\mu} S dt + 
dw_{a,x,\mu})\,, 
\label{cle} 
\ee 
$\lambda_a, a=1\ldots, 8$ are the Gell-Mann matrices; $dw_{x,\mu,a}$ are 
independent Wiener increments, normalized as 
\be 
\bra dw_{a,x,\mu}(t) dw_{a',x',\mu'}(t')\ket= 2\delta_{aa'}\delta_{tt'} 
\delta_{xx'}\delta_{\mu\mu'}dt\,; 
\ee 
$D_{a,x,\mu}$ is a differential operator (derivation) acting on functions 
$f$ of $SU(3)^{\#{\rm links}}$ as
\be 
D_{a,x,\mu} f(\{U\})= \lim_{\delta\to 0} 
\frac{1}{\delta}\left [ f(\{U(\delta)\}) -f(\{U\})\right]\,, 
\label{deriv} 
\ee 
where $\{U(\delta)\}$ means the variable $U_{x,\mu}$ has been replaced by 
$\exp(i\delta \lambda_a)U_{x,\mu}$ with all other variables unchanged. The 
first term on the rhs of Eq.(\ref{cle}) (the drift term) is gauge  
covariant and transverse to the gauge orbits since $S$ is gauge invariant, 
but the noise term contains components along the gauge orbits. Since the 
gauge group is noncompact, the process (1) may go far from the unitary 
submanifold.

An Euler discretization of Eq.(\ref{cle}) (see also \cite{aj}) gives
\be
U_{x,\mu}\mapsto \exp\left\{-\sum_a i\lambda_a (\epsilon 
K_{a,x,\mu}+\sqrt{\epsilon}\eta_{a,x,\mu})\right\}U_{x,\mu}\,,
\label{dyn}
\ee
where $K_{a,x,\mu}= D_{a,x,\mu}S $ is the drift force and $\eta$ are 
independent Gaussian noises satisfying
\be
\bra \eta_{a,x,\mu} \eta_{a',x',\mu'}\ket =2 \delta_{aa'} \delta_{xx'} 
\delta_{\mu\mu'}\,.
\ee 

{\it Gauge cooling.} -- It is a well known problem of the CLE method 
that the system may drift too far out into the imaginary directions, 
causing numerical problems. To quantify the distance from unitarity we 
use the `unitarity norm'
\be
F(\{U\})\equiv \sum_{x,\mu}\tr\left[\, U^\dagger_{x,\mu} U_{x,\mu}
+(U_{x,\mu}^\dagger)^{-1}U_{x,\mu}^{-1}- 2\right]\ge 0\,,
\ee
which vanishes if and only if all the $U$'s are unitary. 

Gauge invariance is the freedom to transform variables 
\be
U_{x,\hat\mu} \mapsto V_x^{-1} U_{x,\mu} V_{x+\hat\mu} 
\ee
for all links $x,x+\hat\mu$, where initially $V_x, V_{x+\hat\mu}$ are 
unitary, but after complexification $V_x, V_{x+\hat\mu}\in SL(3,\C)$. We 
use this freedom to minimize the unitarity norm, without changing the 
observables. Concretely, we intersperse standard `dynamical' Langevin 
sweeps with several `g.c. sweeps'. These are deterministic moves 
essentially in the direction opposite to the gradient of $F$.

A maximally nonunitary gauge transformation at site $y$ in direction $a$ 
is given by 
\begin{align}
U_{y,\mu}&\mapsto \exp(\tilde\alpha\lambda_a)U_{y,\mu}\,\notag\\ 
U_{y-\hat\mu,\mu}&\mapsto U_{y-\hat\mu,\mu} 
\exp(-\tilde\alpha\lambda_a)\,   
\label{mx}
\end{align}
($\tilde\alpha\in\R$) for all $\mu$, with all other link variables  
remaining unchanged. The `gauge gradient' of $F$ in the $a$ direction at 
the lattice site $y$ is then
\begin{align}
G_{a,y}&\equiv D_{a,y}F= 2\tr \lambda_a \left[U_{y,\mu} U^\dagger_{y,\mu}- 
U_{y-\hat\mu,\mu}^\dagger U_{y-\hat\mu,\mu}\right]\notag \\
&+2 \tr \lambda_a \left[-(U_{y,\mu}^\dagger)^{-1}U_{y,\mu}^{-1}
+(U_{y-\hat\mu,\mu}^\dagger)^{-1}U_{y-\hat\mu,\mu}^{-1}\right]\,.
\end{align}
The g.c. updates of the configuration are given by 
\begin{align}
U_{x,\hat\mu}&\mapsto \exp\left(-\sum_a \tilde\alpha \lambda_a 
G_{a,x}\right) U_{x,\hat\mu}\,\notag \\ 
U_{x-\hat\mu,\mu} &\mapsto U_{x-\hat\mu,\mu} \exp\left(\sum_a \tilde\alpha 
\lambda_a G_{a,x}\right), 
\label{gc}
\end{align}
where $\tilde\alpha=\epsilon \alpha$; $\alpha$  determines the strength of 
the g.c. force, whereas $\epsilon$ is a discretization parameter 
as in Eq.(\ref{dyn}). Note that even if $\tilde\alpha$ is not small, 
Eq.(\ref{gc}) is still a gauge transformation; it just might not be 
optimal for reducing $F$. 

\begin{figure}
  \includegraphics[width= 0.9\columnwidth]{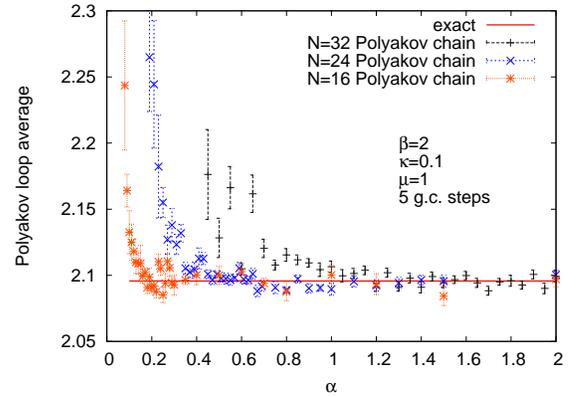}
  \caption{Polyakov chain: average Polyakov loop vs. $\alpha$.}
\label{gf}
\end{figure}

\begin{figure}
  \includegraphics[width= 0.9\columnwidth]{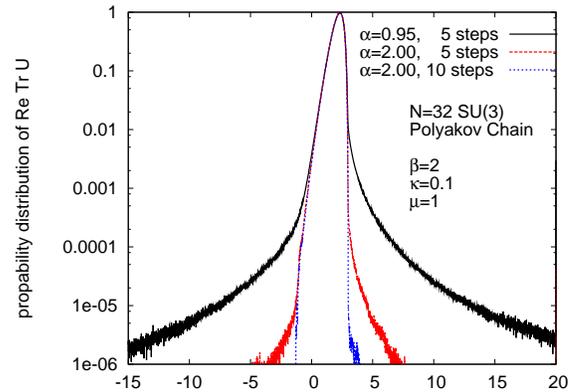}
  \caption{Polyakov chain: histograms of Polyakov loops.} 
\label{histo} 
\end{figure}

{\it Polyakov loop model.} --
The Polyakov loop model is given by a 1D lattice consisting of $N$ links 
with periodic boundary conditions. Analytically it reduces to a one-link 
integral, but it is a useful laboratory to check the effect of g.c.. The 
action is given by
\be
-S = \beta_1 \tr U_1\ldots U_{N_t} +\beta_2 \tr U^{-1}_{N}\ldots 
U^{-1}_1\,,
\ee
where we allow $\beta_{1,2}$ to be complex. Here we choose 
$\beta_1=\beta+\kappa e^\mu$, $ \beta_2= \beta^*+\kappa e^{-\mu}$; $S$ 
will in general be complex. The observables of interest are $\tr\, 
\cP^k=\tr (U_1 \ldots U_{N})^k$, $k=\pm 1,\pm 2,\pm 3$. The effect of the 
g.c.~on $\bra \tr\,\cP\ket$ is shown in Fig.\ref{gf}, where we vary 
$\alpha$ of Eq.(\ref{gc}). In Fig. \ref{histo} we show the effect of 
varying $\alpha$ on the distribution of the values of $\tr\,\cP$. We see 
that with too small $\alpha$ the distribution has a `skirt' of slow 
decay. From \cite{ajss} we know that slow decay typically leads to 
incorrect results; a small skirt, however, does not lead to appreciable 
deviations from the exact values. The main point is that enough cooling 
leads to correct results, see Fig.\ref{gf}.

We also measured $\tr\, U^{\pm 2}$ and $\tr\, U^{\pm 3}$ with similar 
results, for $N$ up to $1024$; it turns out that increasing $N$ requires 
an increase in cooling.

{\it Heavy quark QCD (HQCD).} -- The model was first explored using the 
CLE without g.c. in \cite{aartsstam} on small lattices and for 
moderate $\mu$. It is defined by dropping all spatial hopping terms of 
the quarks; this amounts to simplifying the Wilson fermion determinant of 
lattice QCD with nonzero chemical potential $\mu$ to
\be
\det {\bf M}(\mu)\equiv \, \prod_{{x}}~
 {\rm Det} \left(\one~+\,C {\cP}_{x}\right)^2 
 {\rm Det} \left(\one~+\,C' {\cP_x^{-1}}\right)^2 \,,
\label{hddet}
\ee
where $C=[2\kappa \exp(\mu))]^{N_t}$, $C'=[2\kappa \exp(-\mu)]^{N_t}$, Det
refers to the color degrees of freedom and
\be
\cP_x = \prod_{\tau=0} ^{N_\tau-1}U_{x+r\hat 0,0} \,.
\ee
HQCD is described by the action 
\be 
S=\frac{\beta}{6} S_G(\left\{U\right\})+ \ln 
\det {\bf M}(\mu) 
\label{hqm} 
\ee 
where $S_G=\sum_P {\rm Re}\; \tr(U_P)$ is the Wilson plaquette action 
and antiperiodic b.c. in time are chosen for the fermions. 
We have
\be
{\rm Det} \left(\one~+\,C {\cP}_{x}\right)^2= (1+ C^3+
 3C P_x + 3C^2 P'_x)^2\,.
\label{det}
\ee
and similarly for the second factor in Eq.(\ref{hddet}), where
\be
P_x=\frac{1}{3} \tr\, \cP_x;\quad P_x'=\frac{1}{3} \tr\, \cP^{-1}_x\,.
\ee
A related model has been studied numerically by a reweighting (RW) 
technique to deal with the complex density in \cite{feo} (and more 
recently in \cite{owe}), where for larger $\mu$ and larger lattices the 
sign problem causes bad signal to noise ratios. See \cite{feo} for the 
general setting.

We simulate the model by the CLE on various lattices, with $\kappa=0.12$, 
various values of $\beta$ and $\mu$, using Eq.(\ref{dyn}) with $\epsilon 
= 10^{-5} - 2\times 10^{-4}$. In some cases we use adaptive control for 
the dynamical stepsize and the number of cooling steps. The system is 
thermalized after a cold start up to Langevin time $t=10$ before taking 
averages.

\begin{figure}
  \includegraphics[width= 0.9\columnwidth]{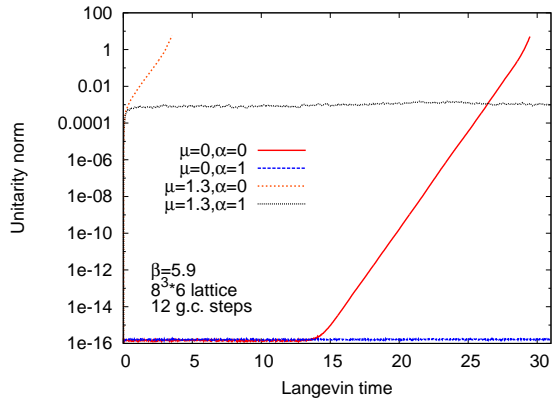}
  \caption{HQCD: evolution of the unitarity norm with and without g.c. for 
     $\beta=5.9$, $8^3\times 6$ lattice.}
\label{unit}
\end{figure}

In Fig.\ref{unit} we show the Langevin evolution of the unitarity norm: it 
is seen how g.c. stabilizes the process. We see that g.c. is needed even 
for $\mu=0$ where the process is real, but has an instability pushing it 
away from the unitary submanifold in the absence of cooling.

\begin{figure}
  \includegraphics[width= 0.9\columnwidth]{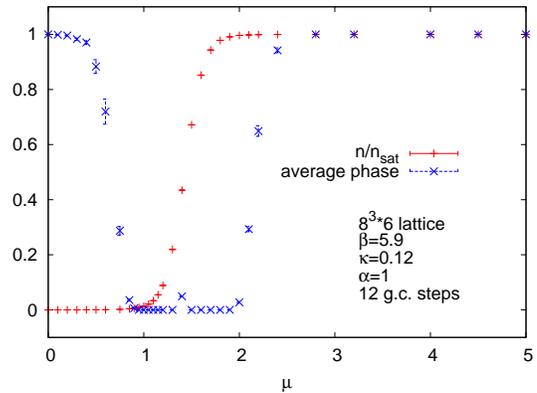}  
  \caption{HQCD: Baryon density and average phase for $\beta=5.9$, 
   $8^3\times 6$ lattice.}
\label{signdens}
\end{figure}

In Fig.\ref{signdens} we show the baryon density $n/n_{sat}$ as a function 
of the chemical potential $\mu$ on an $8^3\times 6$ lattice. We find 
saturation for $\mu \gtrsim 2$, suggesting that our results are correct 
also for fairly large $\mu$. The figure also shows the average phase of 
$\exp(-S)$, defined by
\be
\left\bra e^{2i\phi} \right\ket \equiv 
\left\bra\frac{\det {\bf M}(\mu)}
{\det{{\bf M}(-\mu)}}\right\ket\,; 
\ee
(see \cite{aartsstam}). With growing density the average phase drops to 
almost zero and rises again in the saturation region.

\begin{figure}
  \includegraphics[width= 0.9\columnwidth]{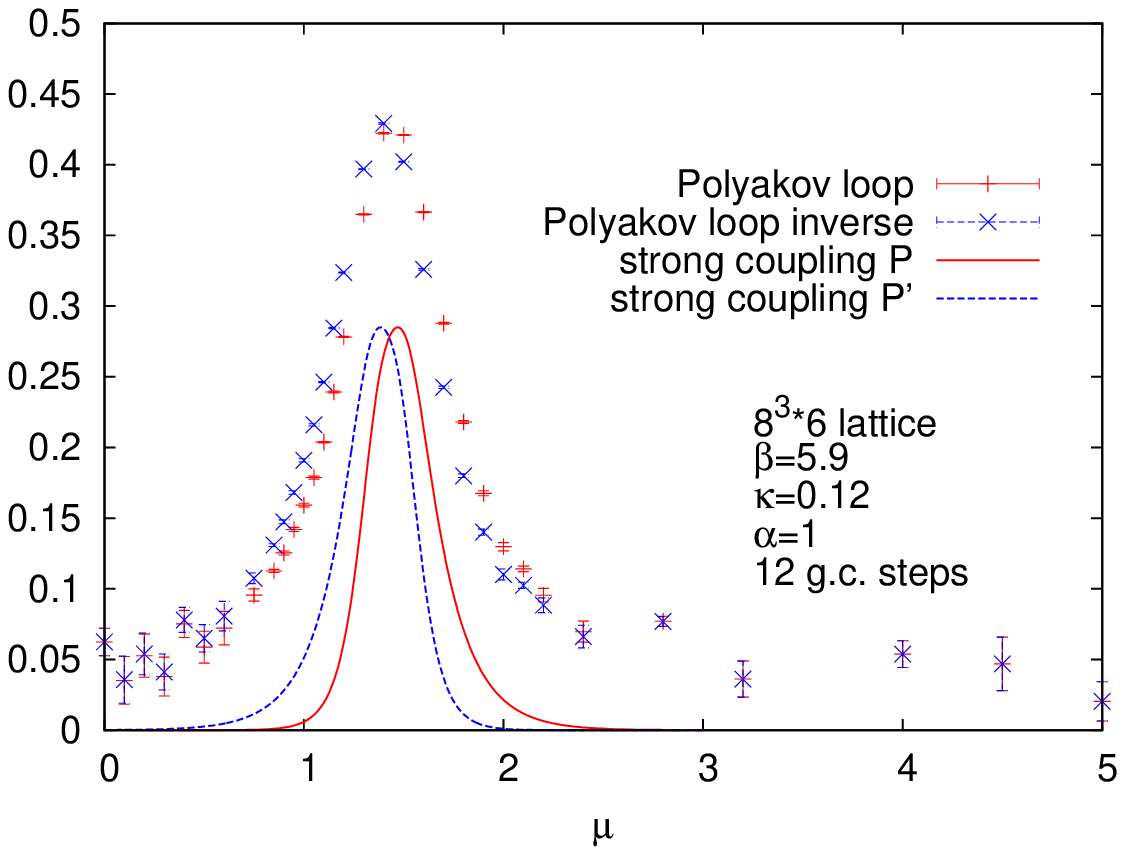}  
  \caption{HQCD: $\bra \cP\ket$, $\bra \cP'\ket$  vs. $\mu$ at 
  $\beta=5.9$ on a $8^3\times 6$ lattice; solid lines: analytic strong 
   coupling result.}
\label{polyakov}
\end{figure}

Fig.\ref{polyakov} shows $\bra P\ket$ and $\bra P'\ket$ vs. $\mu$. Both 
reach a maximum where the density takes off ($\mu\approx 1.4$), then drop 
again to nearly zero (see also \cite{joyce,amaas,owesilver}); the curve 
for $\bra P'\ket$ is shifted slightly to the left with respect to $\bra 
P\ket$. The analytic curves for $\beta=0$ in the figure exhibit a similar 
behavior; they are calculated essentially as in \cite{feo}.

This behavior is due to the fact that for small as well as large $\mu$ the 
determinants in Eqs.~(\ref{hddet}) are dominated by the constant terms, 
leading to small expectation values of $P$ and $P'$. In physics terms, to 
get a nonzero answer one has to form a localized colorless bound state 
between the dynamical quarks and the external charge provided by $P$ or 
$P'$. For small $\mu>0$, when typically there is only 1 quark present, it 
can form a `meson' with $P'$ but not with $P$;  for larger $\mu$, when 
typically 2 quarks are present, $P$, but not $P'$ can form a `baryon' with 
them. The situation is slightly complicated by the fact that up to 6 
quarks can exist at every site, so $P'$ can also form a colorless state 
with 4 and $P$ with 5 quarks. With no quarks or in the completely filled 
state it is not possible to form such bound states. The simple picture 
given here ignores fluctuations, including a small probability, damped by 
$\mu$, of finding antiquarks, so at $\mu=0$ $\bra P\ket $ and $\bra P'\ket 
$ are small but nonzero, whereas at very large $\mu$ they will go to zero. 
Thus both quantities will have a peak, but $\bra P\ket$ will peak later 
than $\bra P'\ket$.

\begin{figure}
  \includegraphics[width= 0.9\columnwidth]{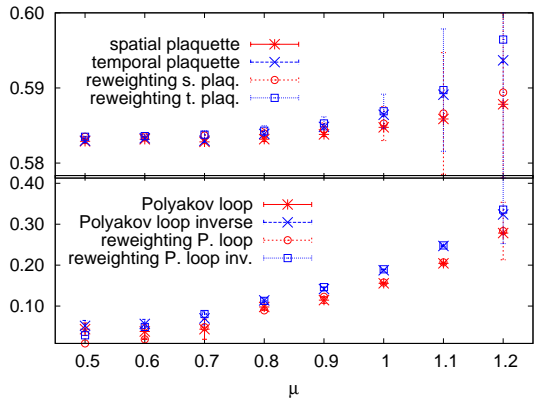}
  \caption{HQCD: Comparison of $\bra P\ket$, $\bra P'\ket$ and plaquettes 
   for  RW and CLE at $\beta=5.9$, $6^4$ lattice, $\alpha=1$, 12 
   g.c. steps.}
\label{comp}   
\end{figure}

{\it Comparison of CLE with RW.} -- To check the CLE results we also 
simulated HQCD with a RW technique, in the region where this is feasible 
(cf.~\cite{feo}). We compare the results for the Polyakov loops and for 
plaquette expectations in Fig.\ref{comp}. The results agree within the 
errors (large errors only occur for RW). Note that around $\mu=1.2$ the RW 
method breaks down, whereas CLE with g.c. works fine for arbitrarily large 
$\mu$.

\begin{figure}
  \includegraphics[width= 0.9\columnwidth]{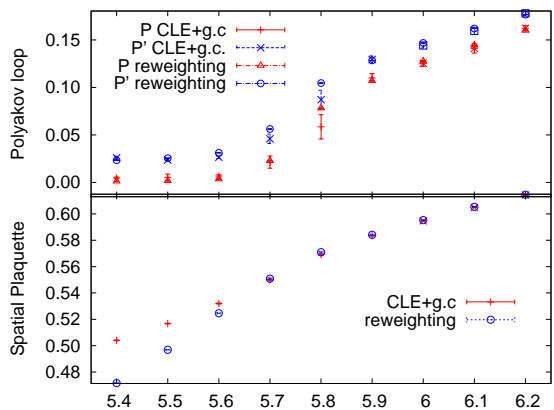}
  \caption{HQCD: $\bra P\ket$, $\bra P'\ket$  and plaquettes 
   vs. $\beta$, $\mu=0.85$ on a $6^4$ lattice with adaptive step size.}
\label{trans}
\end{figure}

In Fig.\ref{trans} we show the expectation values of Polyakov loops at
$\mu=0.85$ vs.~$\beta$, both obtained from RW and CLE. The figure
shows that CLE works for large $\beta$ and also allows us to cross over
into the confining region. Deeper in the confining region, however, we see
small differences between the RW and CLE data, apparently related to an 
emergence of a `skirt' of the distribution.

{\it Concluding remarks.} -- The instabilities that arise without or with 
too weak g.c. are apparently caused by (1) the existence of repulsive 
fixed points, pushing the configurations exponentially fast away from the 
unitary submanifold and (2) roundoff and other numerical errors. Each 
gauge orbit contains configurations arbitrarily far from the unitaries, 
and there roundoff errors pile up to affect also the observables, 
violating 
gauge invariance.

The modified process remedies these problems; it thereby also avoids the 
slow decay of the equilibrium distribution. The CLE method with g.c. is 
tested successfully for the Polyakov loop model. It allows, apparently for 
the first time, to simulate a real QCD like gauge model at finite density 
all the way into the saturation region. The method does not suffer from 
any sign or overlap problem. There is no fundamental obstacle to extend 
the method to full QCD.

{\it Acknowledgments.} -- We are grateful to Gert Aarts for discussions 
and collaboration in the early stage of the work; we also thank Frank 
James and Joyce Myers for useful discussions. I.-O.~S. and E.~S. were 
partly supported by Deutsche Forschungsgemeinschaft; I.-O.~S. 
also thanks the Werner-Heisenberg-Institut for hospitality. A 
large part of the numerical calculations for this project was done on the 
bwGRiD (http://www.bw-grid.de), member of the German D-Grid initiative, 
funded by BMBF and MWFK Baden-W\"urttemberg.

\end{document}